\documentclass[]{raa}            

\usepackage{graphicx,times,multirow,lscape}             

\begin{document}

   \title{The Connection between Radio and $\gamma$-Ray Emission in Fermi/LAT Blazars}

   \volnopage{Vol.0 (201x) No.0, 000--000}      
   \setcounter{page}{1}          

   \author{Xu-Liang. Fan
      \inst{1,2,4}
   \and Jin-Ming. Bai
      \inst{1,2}
   \and Hong-Tao. Liu
      \inst{1,2}
   \and Liang. Chen
      \inst{3}
   \and Neng-Hui. Liao
      \inst{1,2,4}
    }
   \institute{National Astronomical Observatories/Yunnan Observatory, Chinese Academy of Sciences, Kunming 650011, China;~~{\it fxl1987@ynao.ac.cn; baijinming@ynao.ac.cn}\\
        \and Key Laboratory for the Structure and Evolution of Celestial Objects, Chinese Academy of Sciences, Kunming 650011, China\\
        \and Key Laboratory for Research in
        Galaxies and Cosmology, Shanghai Astronomical Observatories, Chinese Academy of Sciences, 80 Nandan Road, Shanghai, 200030, China\\
        \and The Graduate School of China Academy of Sciences, Beijing 100049, China\\
   }
   \date{Received~~; accepted~~}

\abstract{We collect the 2LAC and MOJAVE quasi-simultaneous data to
investigate the radio-$\gamma$ connection of blazars. The cross
sample contains 166 sources. The statistic analysis based on this sample confirms positive
correlations between these two bands, but the correlations become
weaker as the $\gamma$-ray energy increases. The statistic results between various parameters show negative correlations of $\gamma$-ray photon spectral index
with $\gamma$-ray loudness for both FSRQs and BL Lacertae objects, positive correlations of $\gamma$-ray variability index with the $\gamma$-ray loudness for
FSRQs, a negative correlation of the $\gamma$-ray
variability index with the $\gamma$-ray photon spectral index for
FSRQs, and negative correlations of $\gamma$-ray photon spectral index
with $\gamma$-ray luminosity for FSRQs. These results suggest that the $\gamma$-ray variability
may be due to changes inside the $\gamma$-ray emission region like the injected power, rather than changes in the photon density of the external radiation fields, and the variability amplitude tends to be larger as the $\gamma$-rays are closer to the high energy
peak of spectral energy distribution. No correlation of
variability index found for BL Lacertae objects implies that
variability behavior may differ below and above the peak energy.  }

   \authorrunning{Fan, Bai, Liu, Chen \& Liao}
   \titlerunning{Radio-$\gamma$ Connection of Fermi Blazars}

   \maketitle

\section{Introduction}
\label{sect:intro} Blazars, the most extreme class of Active
Galactic Nuclei (AGNs), are observed at almost the full
electromagnetic spectrum from radio to $\gamma$-ray band. It
contains two subclasses called BL Lacertae objects (BL Lacs) and
Flat Spectrum Radio Quasars (FSRQs). The radiation from blazars is
believed to be non-thermal emission of relativistic jets with
small view angles to the line of sight (Urry \& Padovani
\cite{Urry95}). The broadband spectral energy distribution (SED)
shows two components in the $\nu-\nu F_{\nu}$ diagram. The lower
energy component that peaks in infrared to X-ray band is believed
to be produced by synchrotron process of relativistic electrons.
The radiation mechanism of higher energy component is more
debatable until now. It is generally believed that the
$\gamma$-ray emission is produced by the inverse Compton (IC)
process of the same electron population that emits the synchrotron
emission (e.g., Maraschi, Ghisellini \&
Cellotti~\cite{Maraschi92}; Sikora, Begelman \&
Rees~\cite{Sikora94}). This indicates a connection between radio
and $\gamma$-ray emission, but the connection could be indirect because the emission regions of radio and $\gamma$-ray may be different due to synchrotron self-absorption. Moreover, some observations found the connection between these two bands, at least during the flare period, e.g., the $\gamma$-ray flares occur after the ejections of new superluminal components (Jorstad et
al.~\cite{Jorstad01}). However, the location of $\gamma$-ray
emission region and the source of seed photons are still uncertain (Marscher et
al.~\cite{Marscher10}; Le{\'o}n-Tavares et al.~\cite{Leon11};
Tavecchio et al.~\cite{Tavecchio10}), and why only a small fraction of
blazars are $\gamma$-ray loud is still an open question (e.g., Kovalev
et al.~\cite{Kovalev09}; Linford et al ~\cite{Linford11}; Sikora
et al.~\cite{Sikora02}).

The connection between radio and $\gamma$-ray emission can help us
constrain the radiation process and the emission region in the jets,
especially for $\gamma$-ray band. This has been considered since
the EGRET era, but no confirmed conclusion had been made due to
the limits of the sensitivity of the telescope and the erratic
sample (Kovalev et al.~\cite{Kovalev09}). Thanks to the high
sensitivity, broad energy range, and large view field of Large
Area Telescope (LAT, Atwood et
al.~\cite{Atwood09}) on board the Fermi Gamma Ray Space Telescope
(Fermi) launched successfully in 2008, much larger sample and more accurate photon
flux and spectrum can be achieved. Many papers investigated the
connection of these two bands, and almost all of them confirmed a
positive correlation in Fermi blazars (Kovalev et
al.~\cite{Kovalev09}; Ghirlanda et
al.~\cite{Ghirlanda10},~\cite{Ghirlanda11}; Linford et al
~\cite{Linford11},~\cite{Linford12}; Ackermann et
al.~\cite{Ackermann11a}). However, these previous results showed
many different behaviors in different subclasses or energy bands.
Moreover, these studies only used non-quasi-simultaneous data, or
only studied the correlation of apparent fluxes which depend on
distance.

After two years survey of LAT, the second LAT AGN catalog (2LAC)
gives a more accurate classification and association of 1121 AGNs
in $\gamma$-rays, and 886 sources in its clean sample (Ackermann
et al.~\cite{Ackermann11b}). Some radio monitor programs are
processing in Fermi era at different frequency, e.g., MOJAVE
(Monitor of Jets in AGN with VLBA Equipment, Lister et
al.~\cite{Lister09}) program at 15GHz, the OVRO (Owens Valley
Radio Observation, Richards et al.~\cite{Richards11}) program at
15GHz, UMRAO (the University of Michigan Radio Variability
Program) at 4.8, 8.0, and 14.5GHz, and so on. These programs give
us a good chance for investigating the connection between radio
and $\gamma$-ray emission with quasi-simultaneous data.

Abdo et al. (\cite{Abdo11b}) obtained the $\gamma$-ray photon
spectral index fitted with a single power-law and the $\gamma$-ray
variability index, which shows the level of $\gamma$-ray
variability. The spectral index is often discussed in many
researches, as it reflects the peak of spectral energy
distribution (SED, e.g., Ghisellini, Maraschi \&
Tavecchio~\cite{Ghisellini09}; Lister et al.~\cite{Lister11}). It
is possible that the variability index is also important for
researching the $\gamma$-ray emission region and the origin of
$\gamma$-ray variability. Ackermann et al. (\cite{Ackermann11a})
has discussed it in $\gamma$-ray range for 2LAC clean sample.

In this paper, we statistically analyze both flux and luminosity correlations with the
quasi-simultaneous data of 2LAC and MOJAVE for the first time. The
data of seven different $\gamma$-ray energy bands are used to
study the dependence of the radio-$\gamma$ connection on $\gamma$-ray energy bands. The
$\gamma$-radio luminosity ratio is defined as a parameter of
$\gamma$-ray loudness to discuss the $\gamma$-ray spectrum and
variability in detail (Arshakian et al.~\cite{Arshakian12}; Lister
et al.~\cite{Lister11}). Then we examine the correlations of $\gamma$-ray photon spectral
index and $\gamma$-ray variability index with $\gamma$-radio
luminosity ratio and other parameters in our sample.

In section 2, we describe our sample and data, section 3 is the
methods we use and the results, section 4 is discussion and
summary. In this paper, we use a $\Lambda$CDM cosmology model with
h=0.71, $\Omega_{m}$=0.27, $\Omega_{\Lambda}$=0.73 (Komatsu et
al.~\cite{Komatsu09}). We define the radio spectral index as
$S(\nu)\propto\nu^{-\alpha}$, and the $\gamma$-ray photon spectral
index as $dN/dE\propto E^{-\Gamma}$.

\section{Sample and Data}
\label{sect:sample} We choose the sources of 2LAC clean sample
which have been monitored by the MOJAVE program. For the
quasi-simultaneous data, we require a time interval matching the
Fermi data (from 2008-08-04 to 2010-08-01, Abdo et
al.~\cite{Abdo11b}). Considering the gap of the radio monitor, and
the probable variability time lag in these two bands, we use a
broad time interval for radio data during 2008-01 to 2011-04. In
order to perform the K-correction and luminosity calculations, we
reject the sources without redshift data. Finally, our sample
contains 166 sources, with 127 FSRQs and 39 BL Lacs (hereafter
cross sample). For FSRQs, there are 119 LSPs, and 8 sources without SED classification. For BL Lacs, there are 7 HSPs, 13 ISPs, 18 LSPs, and 1 source without SED classification. The classification and redshift data of our cross
sample are from 2LAC (Ackermann et al.~\cite{Ackermann11b}). The
K-corrections for radio flux and $\gamma$-ray photon flux
are calculated as
\begin{equation}
S_{\nu_{0}}=S_{\nu_{0}}^{'}(1+z)^{\alpha+1} \end{equation}
and
\begin{equation}
N(E_{1},E_{2})=N^{'}(E_{1},E_{2})(1+z)^{\Gamma-1},
\end{equation}
respectively. The radio and $\gamma$-ray luminosity are calculated
by the observed flux as
\begin{equation}
L_{\nu_{0}}=4\times10^{-26}\pi d_{L}^{2}S_{\nu_{0}}^{'}(1+z)^{\alpha-1}~~~~~~~erg~s^{-1}~Hz^{-1}
\end{equation}
and  (Ghisellini, Maraschi \&
Tavecchio~\cite{Ghisellini09})
\begin{equation}
L(\nu_{1},\nu_{2})=4\pi
d_{L}^{2}S^{'}(\nu_{1},\nu_{2})(1+z)^{\Gamma-2}~~~~~~~erg~s^{-1},
\end{equation}
respectively. In equations (3) and (4), $d_{L}$ is the luminosity distance, and $S^{'}(\nu_{1},\nu_{2})$ is the $\gamma$-ray energy flux observed between frequencies $\nu_{1}$ and $\nu_{2}$ which correspond to the energies $E_{1}$ and $E_{2}$. $S^{'}(\nu_{1},\nu_{2})$ is derived as
\begin{equation}
S^{'}(\nu_{1},\nu_{2})=\frac{(1-\Gamma)h\nu_{1}N^{'}(E_{1},E_{2})[(\frac{\nu_{2}}{\nu_{1}})^{2-\Gamma}-1]}{(2-\Gamma)[(\frac{\nu_{2}}{\nu_{1}})^{1-\Gamma}-1]}~~~~~erg~s^{-1}~cm^{-2}~~~\Gamma\neq2.
\end{equation}
Due to the absence of quasi-simultaneous data of radio
spectral index, and generally the flat radio spectrum for blazars, we
assume that the radio spectral index is equal to 0 for the sources in our sample.

For Fermi data, we use the 1GeV-100GeV photon flux in 2LAC
(Ackermann et al.~\cite{Ackermann11b}), and the 100MeV-300MeV,
300MeV-1GeV, 1GeV-3GeV, 3GeV-10GeV data in 2FGL (Abdo et
al.~\cite{Abdo11b}). For upper limits, we use them for our
analysis directly (Ackermann et al.~\cite{Ackermann11a}), but we
remove 10GeV-100GeV data because too many sources only get upper
limits in this band. We obtain the 100MeV-1GeV photon flux
combined with the 100MeV-300MeV and 300MeV-1GeV data, the
100MeV-100GeV photon flux combined with the 100MeV-300MeV,
300MeV-1GeV and 1GeV-100GeV data. For MOJAVE data, we use the mean
value of all the VLBA observations for each source.

\section{Result}
\label{sect:result} We use the nonparametric Spearman test for
our correlation analysis except that the luminosity correlation is
tested by partial correlation analysis to remove the dependence on
redshift. In the paper, we define a significant correlation when the chance probability is less than 0.001.

The data of 15GHz radio flux and the $\gamma$-ray photon flux
(100MeV-100GeV) are plotted in figure 1. The Spearman test confirms
the positive correlation of radio and $\gamma$-ray flux, though the correlation has certain dispersion. This means that the sources with higher radio flux always have higher $\gamma$-ray flux. The correlation for BL Lacs seems to be better than that for
FSRQs. The correlation coefficient and chance probability are 0.38
and 1.1$\times10^{-5}$ for FSRQs, and 0.44 and 0.006 for BL Lacs,
respectively.

Then we examine the correlation between the photon flux of
individual $\gamma$-ray energy band and the radio flux. The
correlations become worse when the energy band increases,
especially for BL Lacs. This indicates that the connection between
radio and $\gamma$-ray emission weakens as the $\gamma$-ray energy
increases. There are no correlations of the radio flux with those
of 1GeV-3GeV, 3GeV-10GeV, and 1GeV-100GeV for BL Lacs. Otherwise, for
other bands including 100MeV-100GeV, the correlation coefficients
of BL Lacs are larger than those of FSRQs. For the radio and
gamma-ray luminosity, we find positive correlations in all energy
bands with the correlation coefficients of BL Lacs larger than
those of FSRQs (Figure 2). Correlation coefficients decrease when
$\gamma$-ray energy gets higher. Figure 3 shows the changes of
correlations with four energy bands, 100MeV-300MeV,
300MeV-1GeV, 1GeV-3GeV and 3GeV-10GeV, respectively.

\begin{figure}[h]
  \begin{minipage}[t]{0.495\linewidth}
  \centering
   \includegraphics[width=80mm,height=60mm]{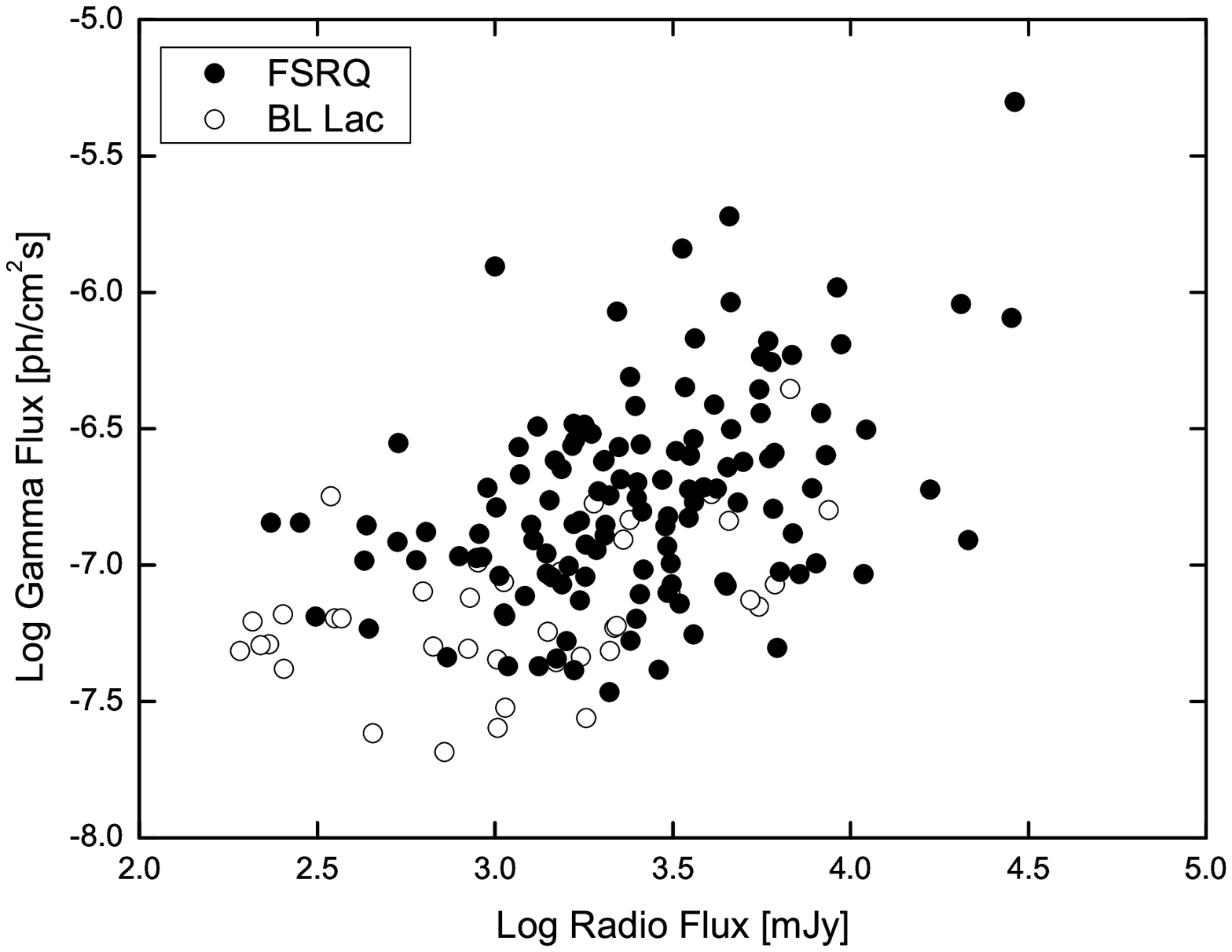}
   \caption{15GHz radio flux versus $\gamma$-ray photon flux (100MeV-100GeV). The filled circles represent FSRQs and the open ones represent BL Lacs.}
  \end{minipage}%
  \begin{minipage}[t]{0.495\textwidth}
  \centering
   \includegraphics[width=80mm,height=60mm]{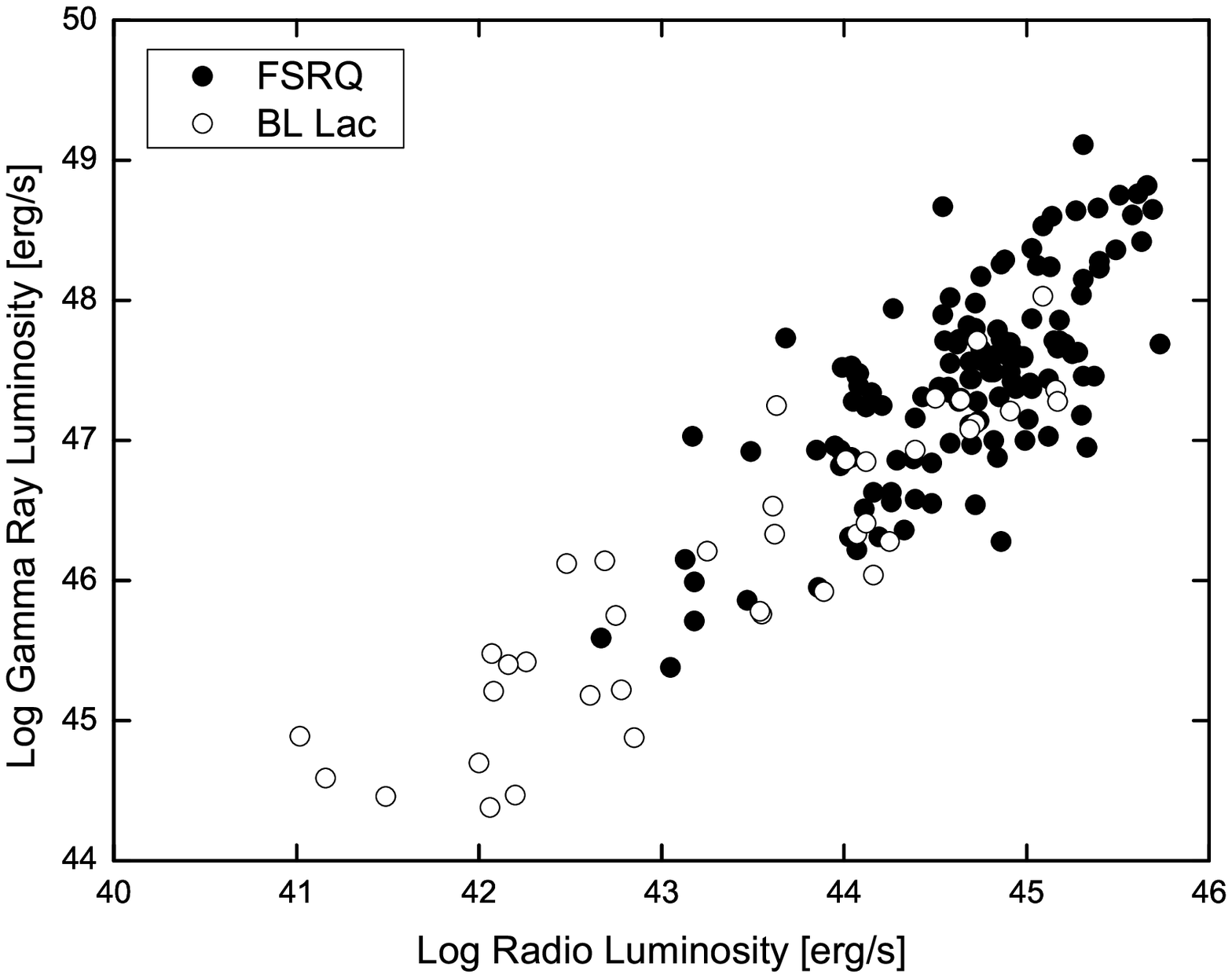}
  \caption{15GHz radio luminosity versus $\gamma$-ray integral luminosity (100MeV-100GeV). The symbols are same as figure 1.}
  \end{minipage}%
  \label{Fig:fig12}
\end{figure}

\begin{figure}[h]
    \includegraphics[width=8cm,angle=0]{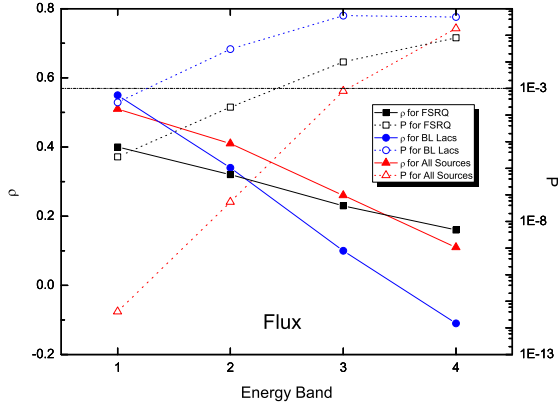},
    \includegraphics[width=8cm,angle=0]{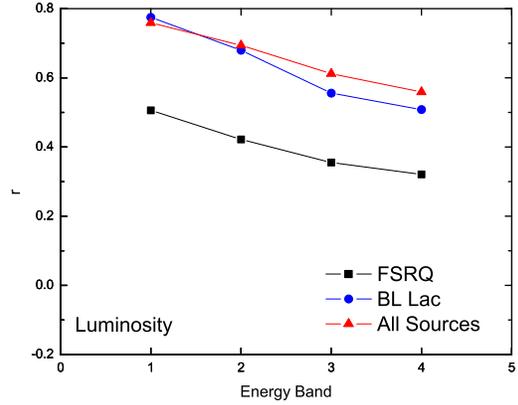}
    \caption{Changes of correlations with energy bands. The black filled squares represent correlation coefficients for FSRQs, the blue filled circles represent correlation coefficients for BL Lacs, the red filled triangles represent correlation coefficients for all sources in our cross sample, while the open ones represent the chance probabilities, respectively. Left shows the change of flux correlations, and the dot-dashed line represents 0.001 of the chance probability. Right shows the change of luminosity partial correlation coefficients. The corresponding chance probabilities are almost all less than 0.001, only one is 0.001.}
   \label{Fig:fig3}
\end{figure}

Figure 4 shows the plot of the $\gamma$-ray photon spectral index
versus the $\gamma$-radio luminosity ratio (100MeV-100GeV). BL
Lacs and FSRQs have a similar range in $\gamma$-ray loudness, and
both of them have negative correlations between the $\gamma$-ray
photon spectral index and the $\gamma$-ray loudness in most of
$\gamma$-ray energy bands, with the correlations for BL Lacs
better than those for FSRQs. For FSRQs, we find positive
correlations between the $\gamma$-radio luminosity ratio and
$\gamma$-ray variability index (Figure 5). The $\gamma$-ray
variability index is negatively correlated with the $\gamma$-ray
photon spectral index for FSRQs (Figure 6).

\begin{figure}[h]
  \begin{minipage}[t]{0.495\linewidth}
  \centering
   \includegraphics[width=80mm,height=60mm]{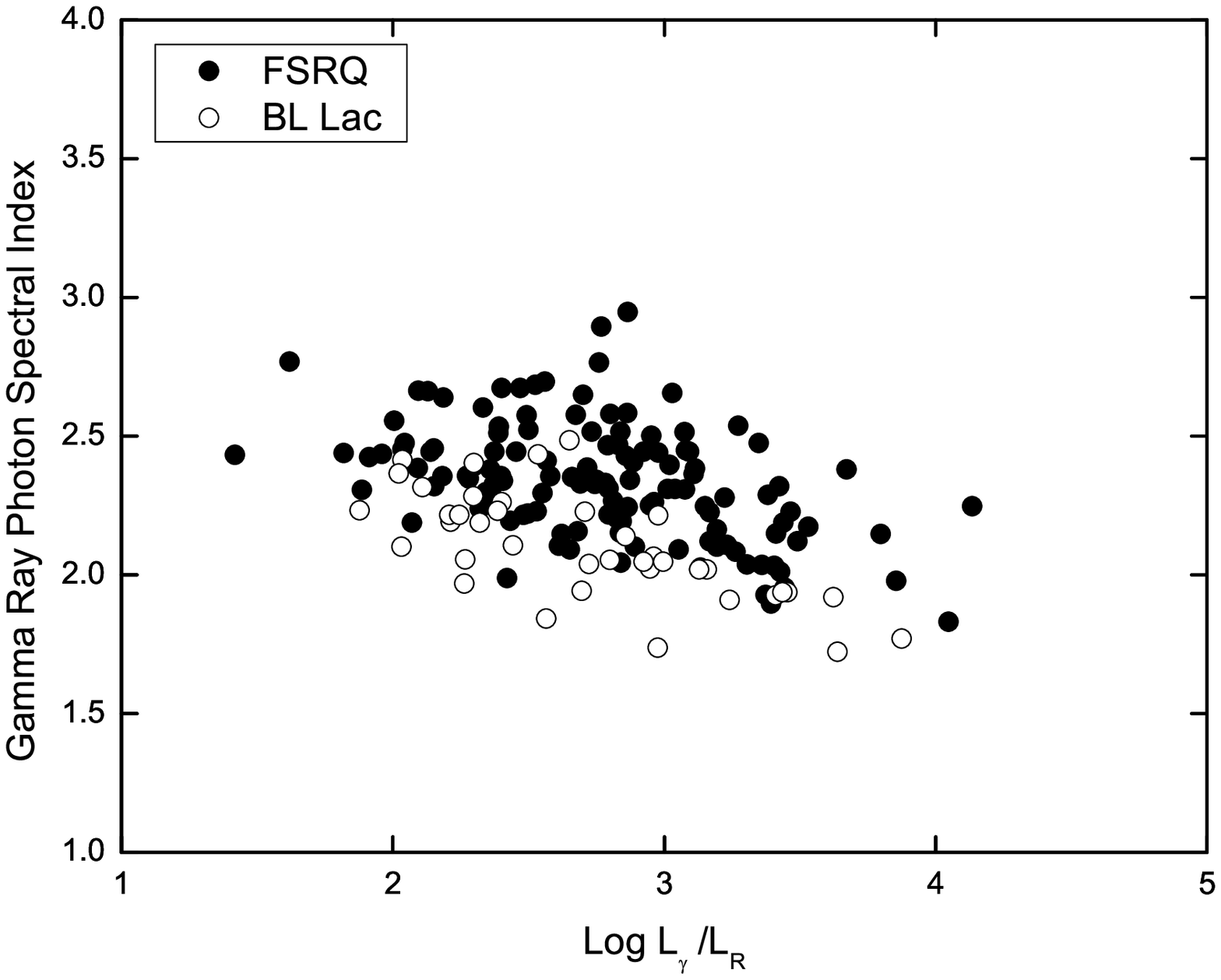}
   \caption{The $\gamma$-radio luminosity ratio ($\gamma$-ray luminosity is from 100MeV to 100GeV) versus $\gamma$-ray photon spectral index. The symbols are same as figure 1.}
  \end{minipage}%
  \begin{minipage}[t]{0.495\textwidth}
  \centering
   \includegraphics[width=80mm,height=60mm]{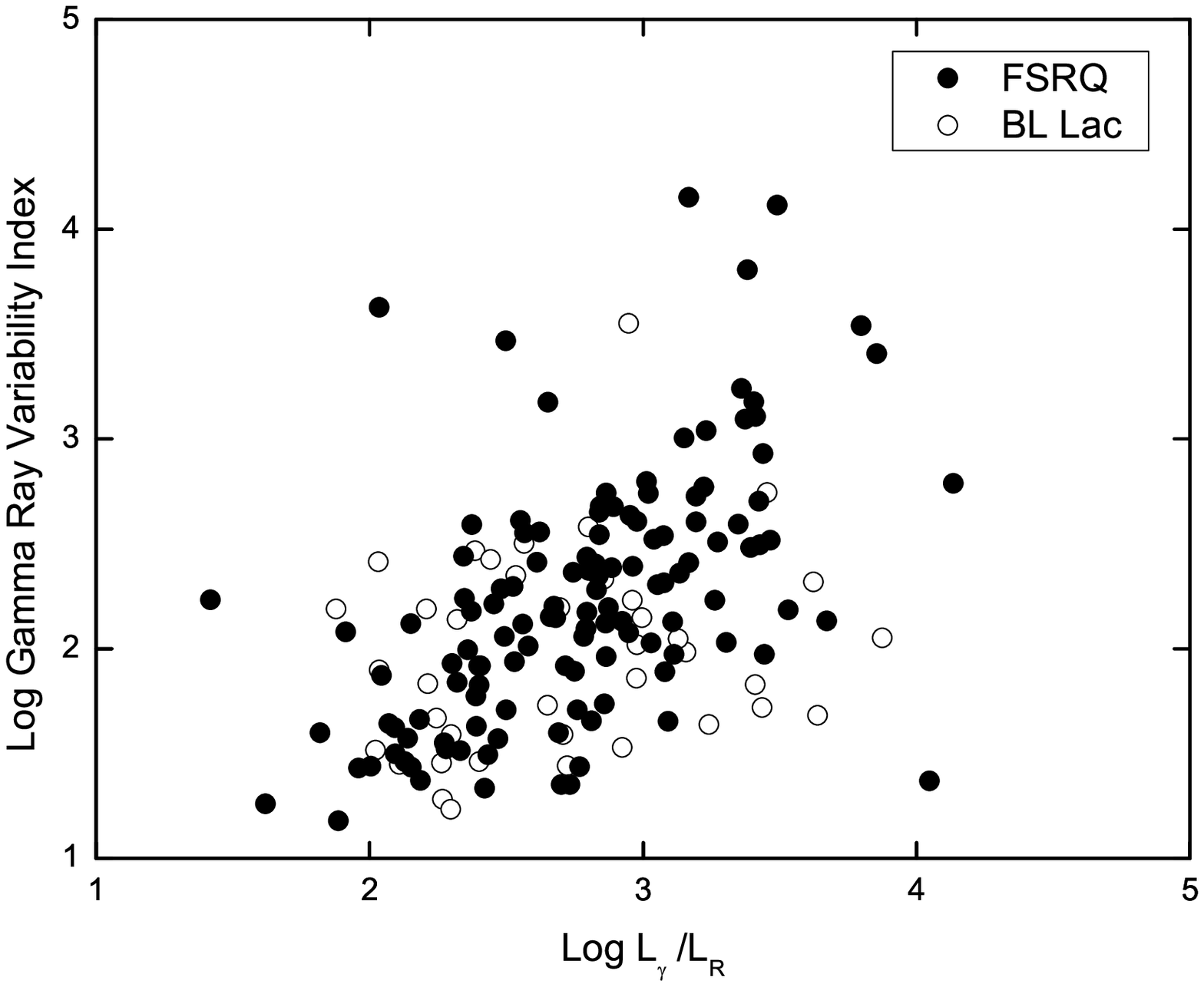}
  \caption{The $\gamma$-radio luminosity ratio ($\gamma$-ray luminosity is from 100MeV to 100GeV) versus $\gamma$-ray variability index. The symbols are same as figure 1.}
  \end{minipage}%
  \label{Fig:fig45}
\end{figure}

\begin{figure}[h]
  \begin{minipage}[t]{0.495\linewidth}
  \centering
   \includegraphics[width=80mm,height=60mm]{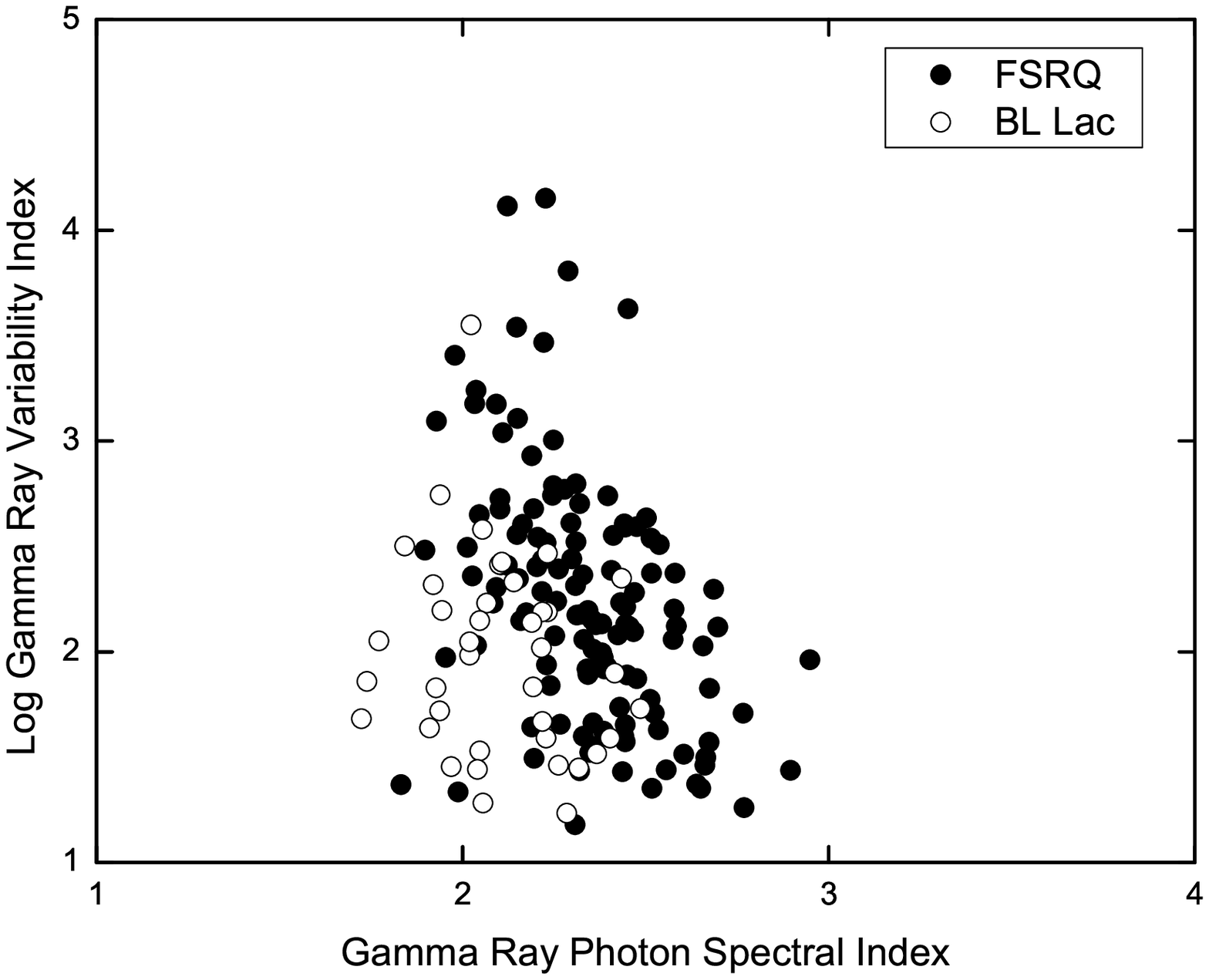}
   \caption{The $\gamma$-ray photon spectral index versus $\gamma$-ray variability index. The symbols are same as figure 1.}
  \end{minipage}%
  \begin{minipage}[t]{0.495\textwidth}
  \centering
   \includegraphics[width=80mm,height=60mm]{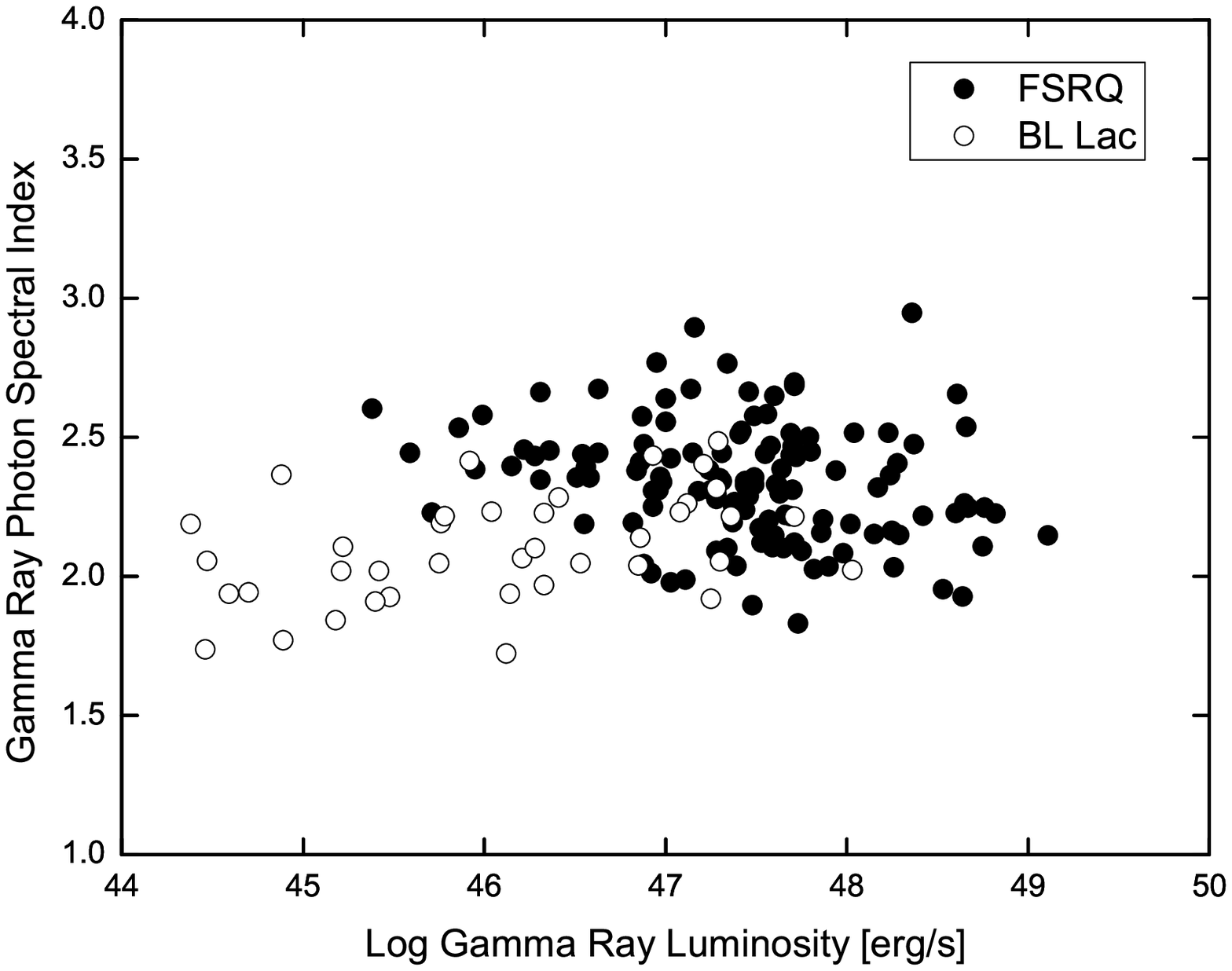}
  \caption{The $\gamma$-ray integral luminosity (100MeV-100GeV) versus $\gamma$-ray photon spectral index. The symbols are same as figure 1.}
  \end{minipage}%
  \label{Fig:fig67}
\end{figure}

We also examine the correlations of radio flux and luminosity
versus $\gamma$-ray photon spectral index and $\gamma$-ray
variability index, respectively. We find the same trends for flux
and luminosity. They have good correlations between radio flux and
luminosity versus $\gamma$-ray photon spectral index for BL Lacs,
respectively. There are no correlations for FSRQs. There are no
correlations between the radio flux and luminosity and the
$\gamma$-ray variability index (see Table 1). For the $\gamma$-ray
luminosity of individual energy band versus the $\gamma$-ray
photon spectral index and the $\gamma$-ray variability index, we
find positive correlations between the $\gamma$-ray
luminosity and the variability index for FSRQs (see Table 1), faint positive
correlations between the $\gamma$-ray luminosity and the photon
spectral index for BL Lacs, and weak negative correlations
between the $\gamma$-ray luminosity and the photon spectral index
for FSRQs, especially in GeV band (see Figure 7).

All the correlation coefficients and chance probabilities are
listed in table 1 except for the case of the radio flux versus the
$\gamma$-ray photon spectral index and the $\gamma$-ray
variability index.

\section{Discussion}
\label{sect:discussion} For our cross sample, we make the Kolmogorov-Smirnov (K-S) tests
of the redshift distribution compared with the 2LAC clean sample
and a full blazar sample, i.e., the bzcat blazar sample (Massaro
et al.~\cite{Massaro09}). There is an intrinsic difference of the
redshift distribution between the cross sample and the bzcat
sample (D = 0.184, probability = 0.0005 for FSRQs, and D = 0.291,
probability = 0.003 for BL Lacs). There is no obvious difference
between 2LAC clean sample and our cross sample (D = 0.045,
probability = 0.991 for FSRQs, and D = 0.159, probability = 0.366
for BL Lacs, see Figure 8). This means that there is an intrinsic difference in redshift distribution between 2LAC clean sample and bzcat sample, which may be caused by the limited number of $\gamma$-ray detected sources, or the special physics mechanisms for $\gamma$-ray sources.

\begin{figure}[h]
    \includegraphics[width=47mm,angle=0]{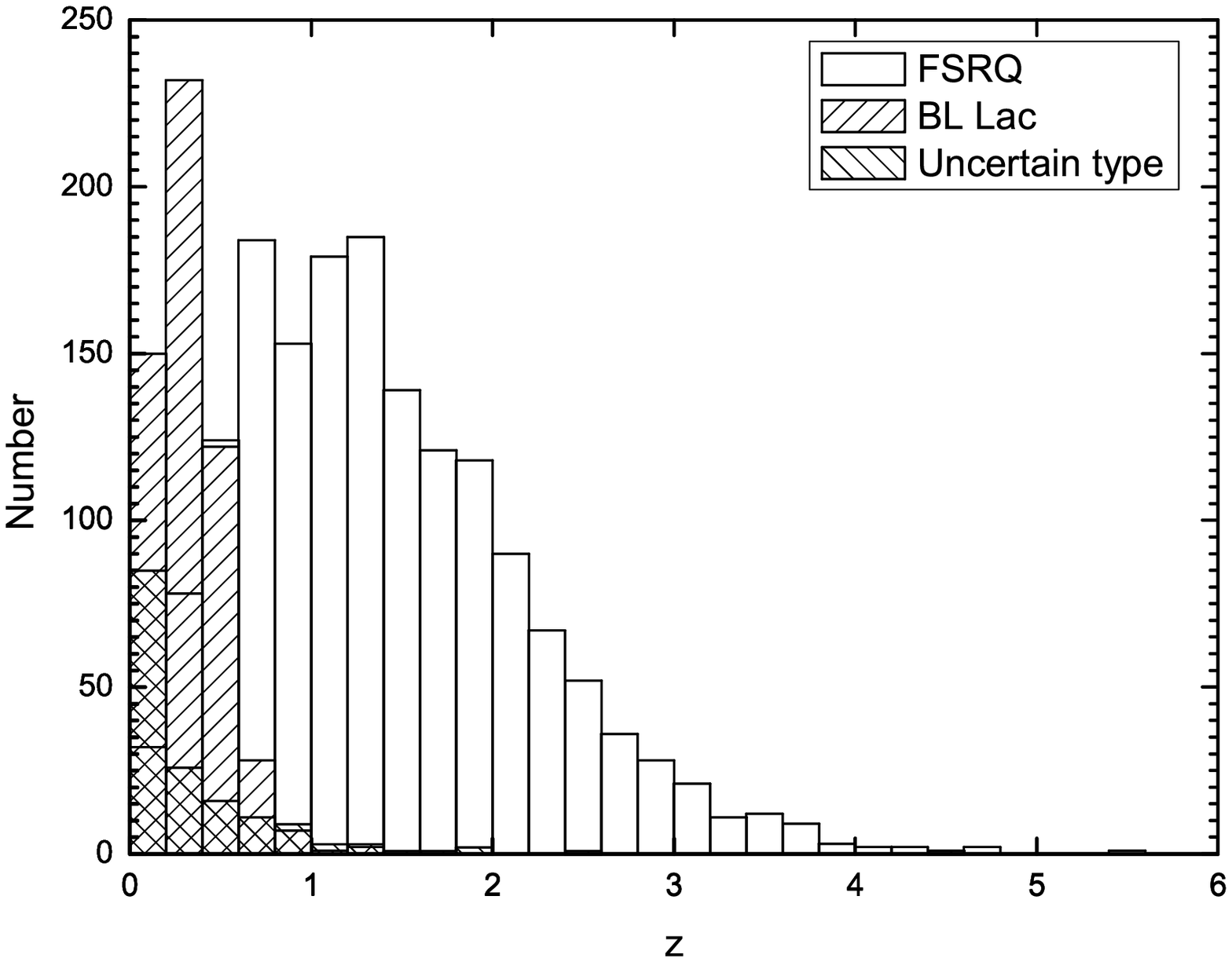},
    \includegraphics[width=47mm,angle=0]{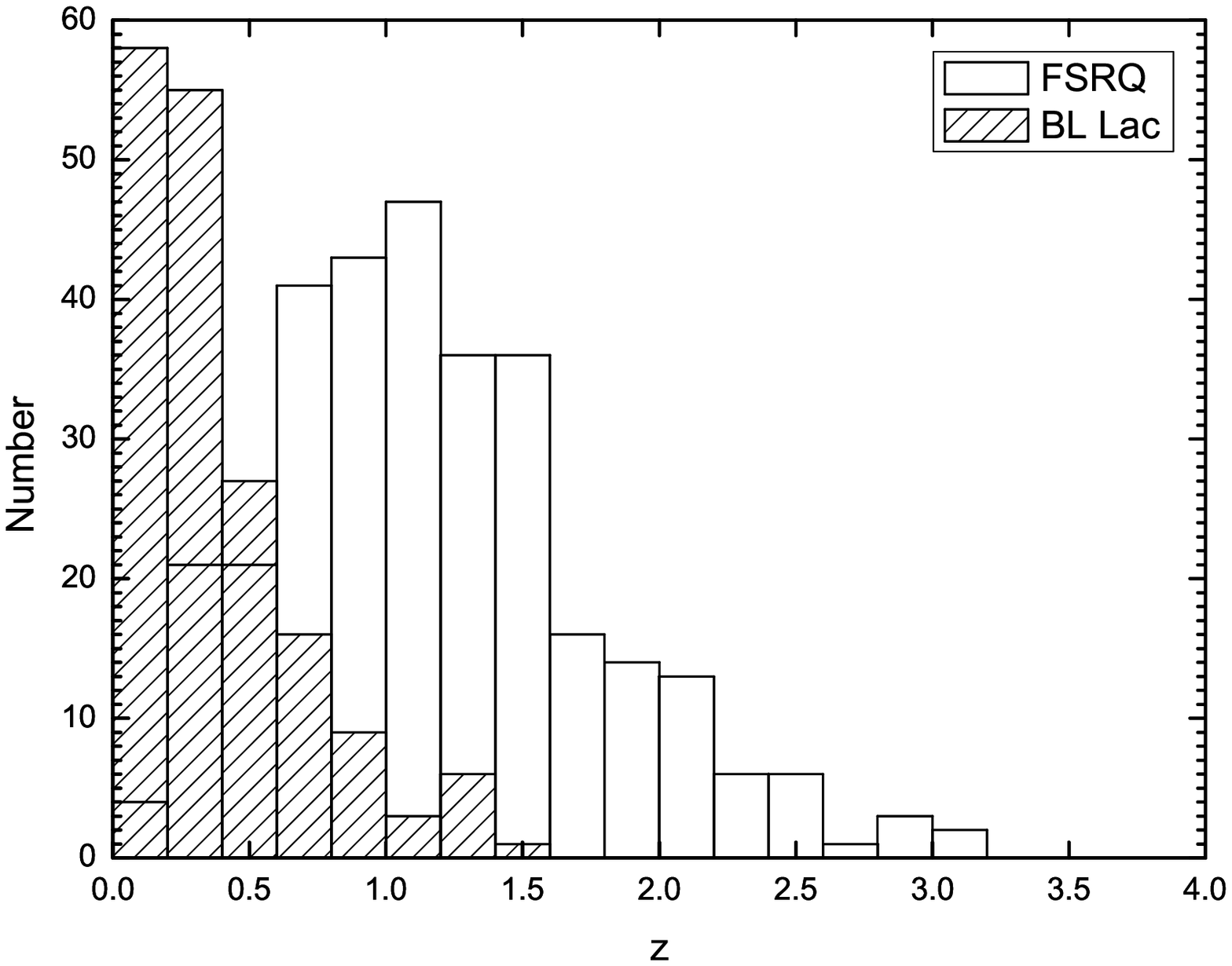},
    \includegraphics[width=47mm,angle=0]{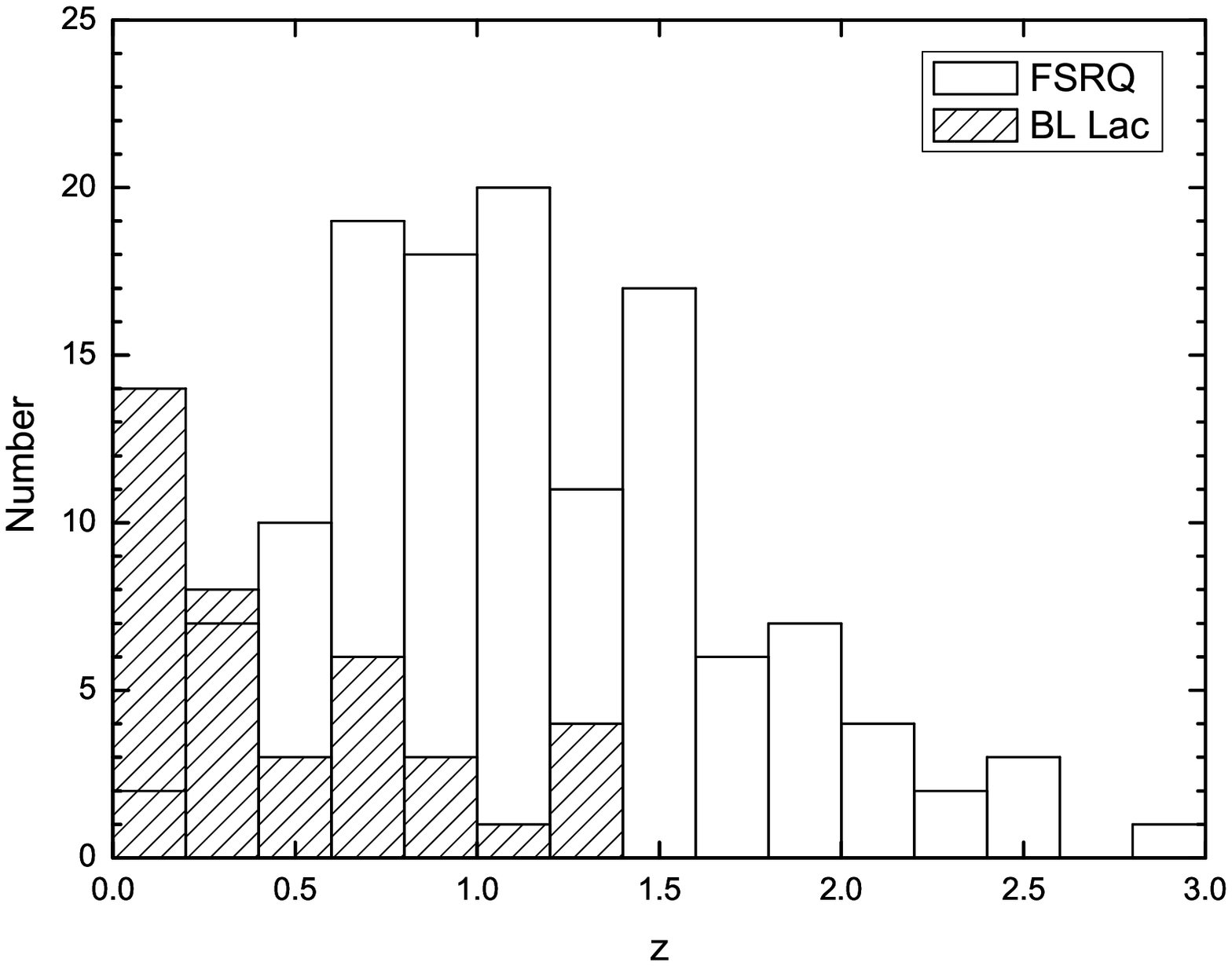}
   \caption{The reshift distributions of three samples. Left is a full blazar sample as bzcat. Center is 2LAC clean sample. Right is our cross sample.}
   \label{Fig:fig8}
 \end{figure}

There is no difference for FSRQs in $\gamma$-ray photon spectral
index distributions of our cross sample and 2LAC clean sample (D =
0.068, probability = 0.78). For BL Lacs, the sources in our sample tend to have soft
$\gamma$-ray spectrum than those in 2LAC clean sample (see Figure 9). This may be caused by a
small fraction of HSP BL Lacs in our sample, because HSPs are generally weaker than ISPs and LSPs at the radio band (Ackermann et al.~\cite{Ackermann11a}). For
all the sources of our sample, they tend to have higher
variability index than the 2LAC sources (Figure 10). This means
that our sample selected the more variable sources and the more
bright sources in $\gamma$-ray band because the variable sources
determined by variability index must be both intrinsically
variable and sufficiently bright (Ackermann et
al.~\cite{Ackermann11b}).

\begin{figure}[h]
    \includegraphics[width=5cm,angle=0]{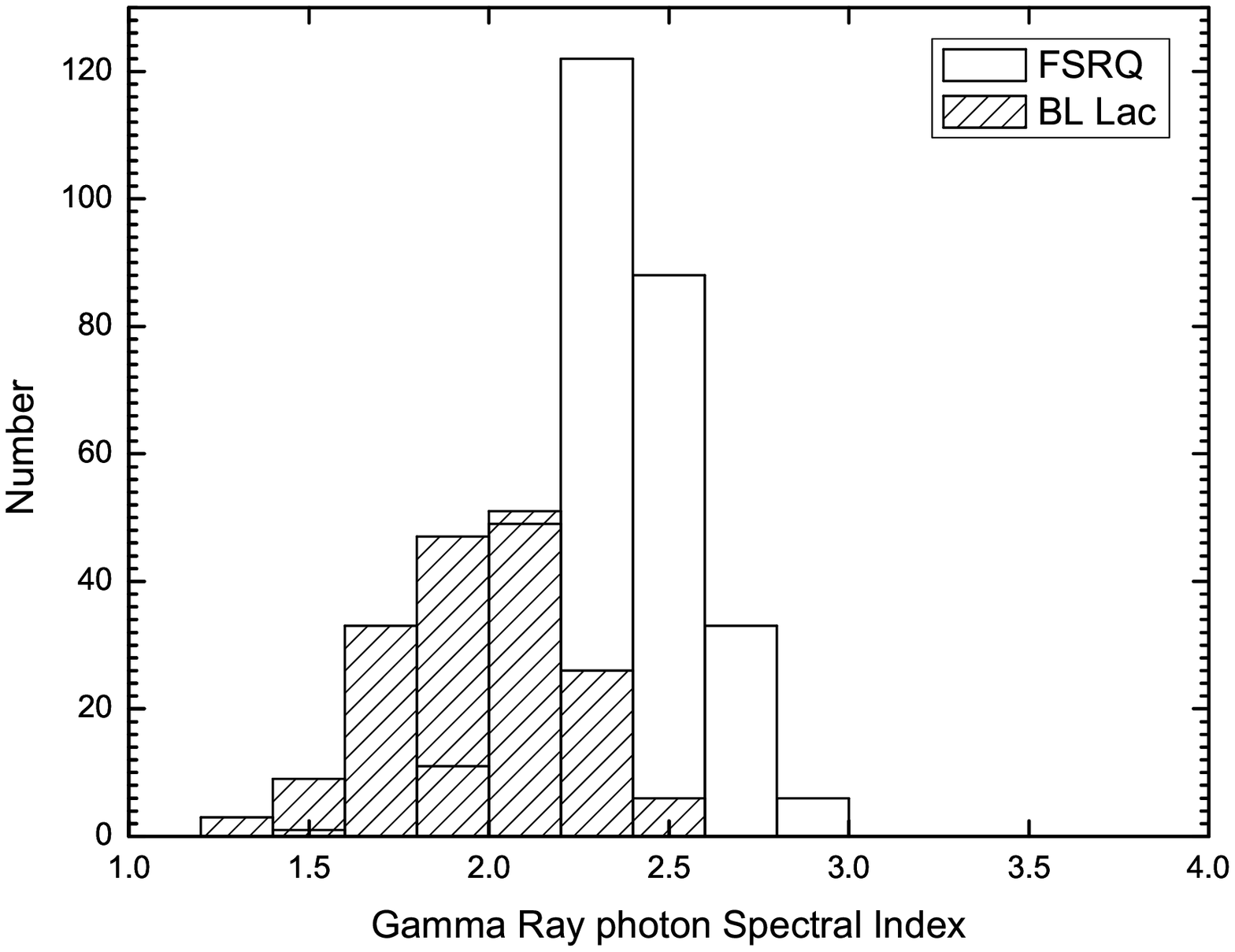},
    \includegraphics[width=5cm,angle=0]{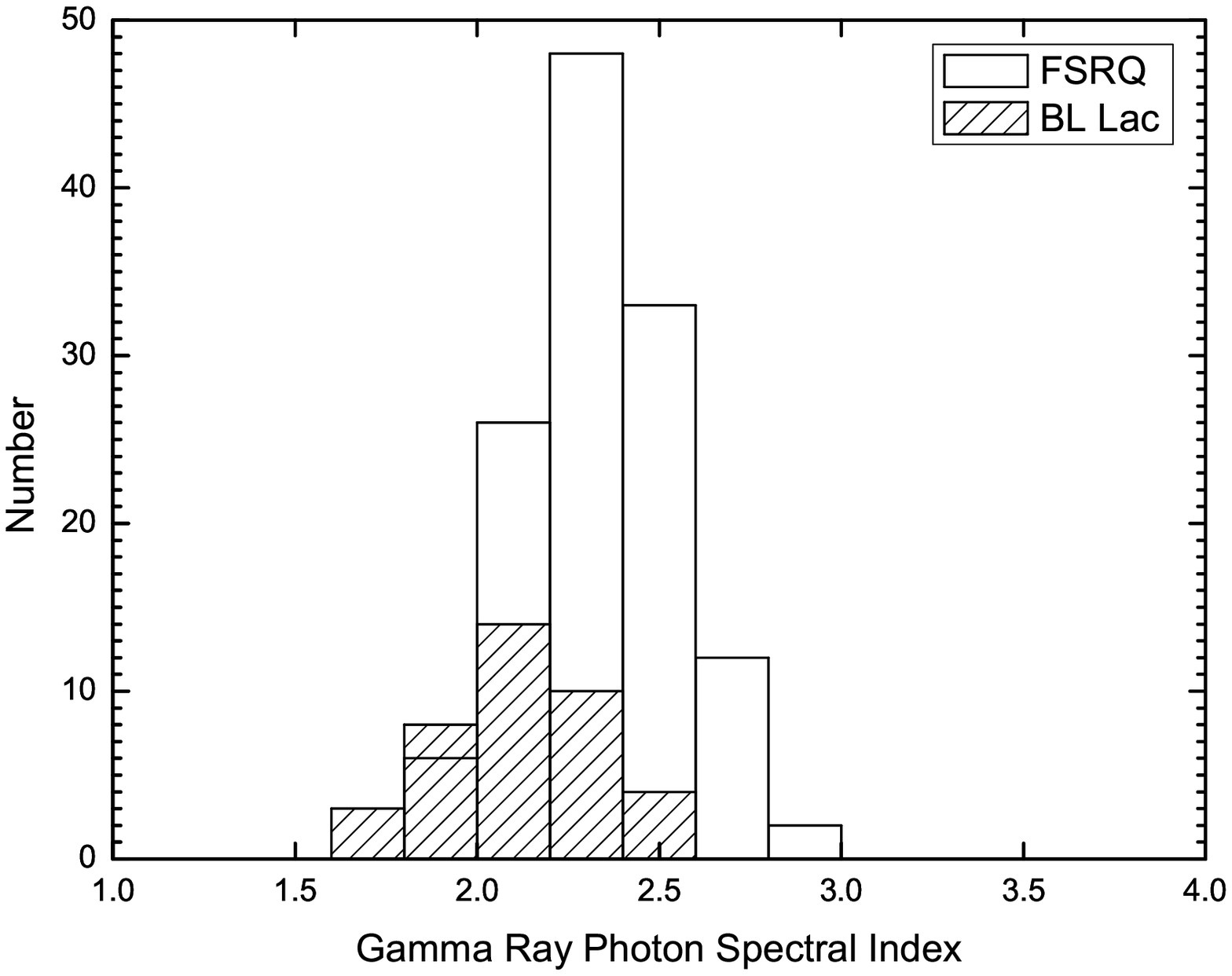}
   \caption{The $\gamma$-ray photon spectral index distributions of two samples. Left is 2LAC clean sample. Right is our cross sample.}
   \label{Fig:fig9}
\end{figure}

\begin{figure}[h]
    \includegraphics[width=5cm,angle=0]{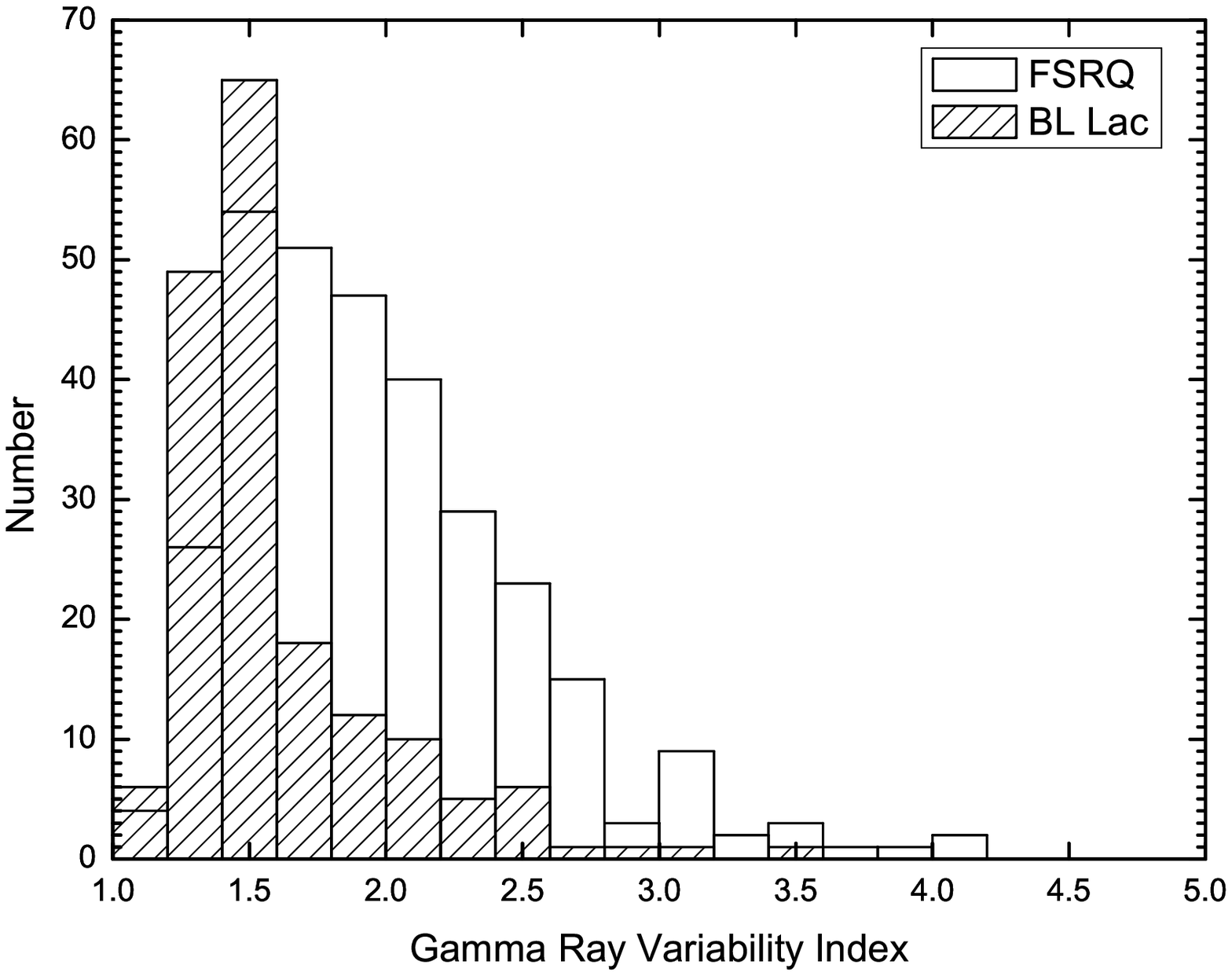},
    \includegraphics[width=5cm,angle=0]{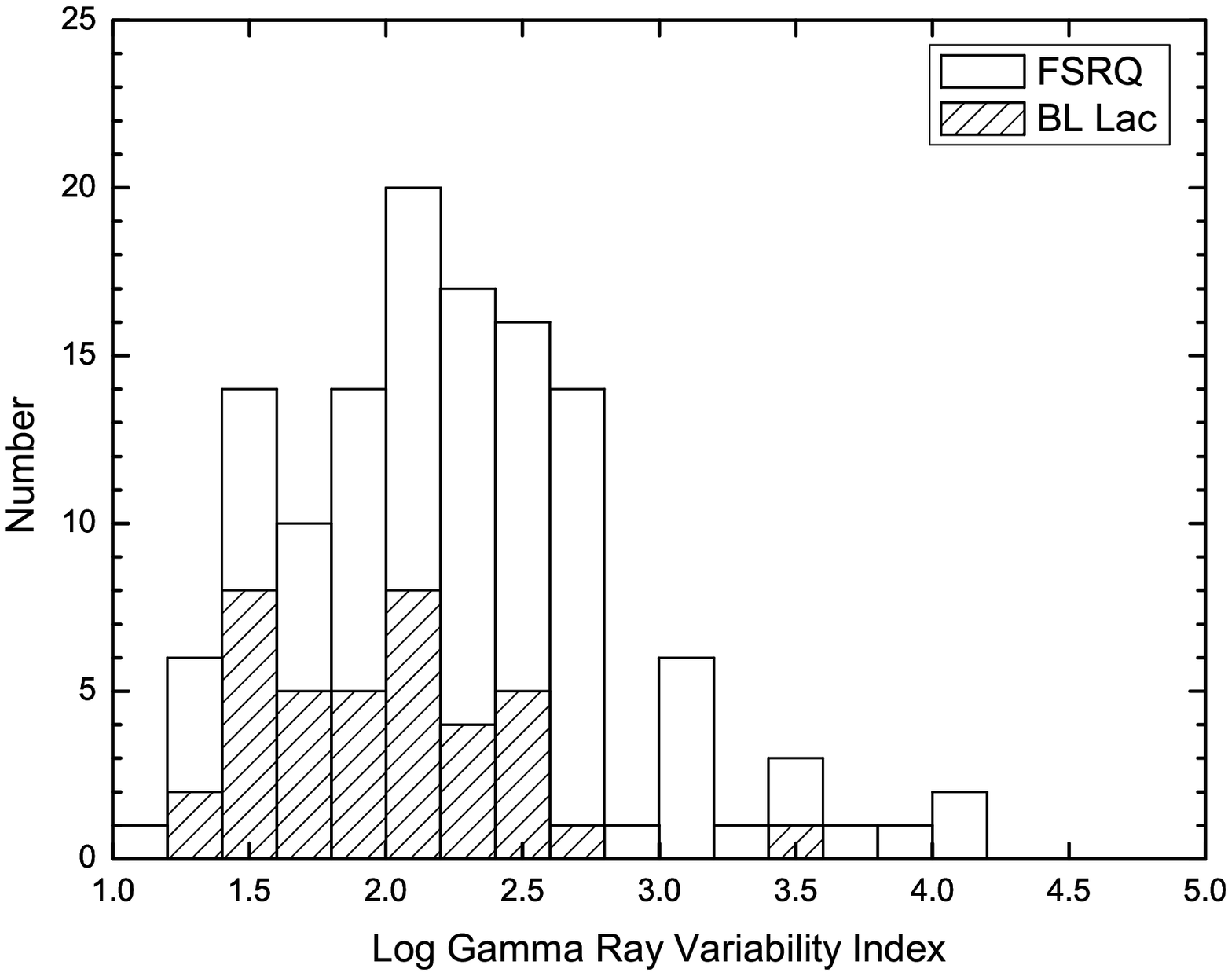}
   \caption{The $\gamma$-ray variability index distributions of two samples. Left is 2LAC clean sample. Right is our cross sample.}
   \label{Fig:fig10}
\end{figure}

Ackermann et al. (\cite{Ackermann11a})
found the different behaviors of different energy bands for flux
correlation. They found that BL Lacs exhibited larger $\rho$
values than FSRQs in the full sample, but a opposite trend in OVRO
sample. In OVRO sample, no correlation were found for BL Lacs above
1GeV, which did not appear in the full sample. The different
results of these three samples may be caused by the selection
effects such as different redshift distributions or different
proportions of HSPs, ISPs, and LSPs in different samples. So we also
remind that the radio-$\gamma$ connection, especially the flux
correlation, should be used very carefully.

The luminosity correlations between radio and individual
$\gamma$-ray energy band are confirmed by our sample. However, both flux and luminosity correlation coefficients decrease as the $\gamma$-ray energy increases. For HSPs, the 15GHz and the GeV emission would be produced by low energy electrons,
and there would be good correlations in these bands. But for ISPs and LSPs, the emission of low energy electrons could only extend to MeV band, and the correlations of radio and GeV bands would be weaker than those of radio and MeV bands. This would be the main
reason why the correlation coefficients decrease as the
$\gamma$-ray energy increases for both BL Lacs and FSRQs.
Moreover, other reasons may also generate these trends, such as,
other emission contents or emission regions in the GeV band.

Lister et al. (\cite{Lister11}) found a correlation between
$\gamma$-ray loudness and $\gamma$-ray photon spectral index for
BL Lacs, but no correlation for FSRQs. In our results, except for
the energy below 1GeV, both BL Lacs and FSRQs have negative
correlations between the $\gamma$-ray loudness and the
$\gamma$-ray photon spectral index. According to the average SEDs
of blazar sequence (Fossati et al.~\cite{Fossati98}), as the
$\gamma$-ray spectrum becomes harder, i.e., the
$\gamma$-ray photon spectral index decreases, the $\gamma$-ray peak of SED
shifts to the higher energy. At the same time, in the energy band of Fermi/LAT, the
ratio of the $\gamma$-ray luminosity to the radio luminosity would
increase significantly. Then it is expected that the $\gamma$-ray loudness is negatively
correlated with the $\gamma$-ray photon spectral index. This expectation is consistent with our
results. This trend is more remarkable among different subclasses
for BL Lacs. For FSRQs, it becomes more remarkable as
the energy is larger than 1GeV (see Table 1). This is due to the effective
cooling above the SED peak.

The correlations between the $\gamma$-ray variability and the
$\gamma$-radio luminosity ratio suggest that the variability, at
least the main variability may originate inside the
$\gamma$-ray emission region like the changes of the injected power, because the changes in photon density of external radiation field such as variability of accretion disk could not affect the $\gamma$-ray loudness. This is opposite to the view of
Paggi et al. (\cite{Paggi11}). These different trends of BL Lacs
and FSRQs imply that variability behavior may distinguish below
and above the peak energy (Ulrich, Maraschi \&
Urry~\cite{Ulrich97}; Ackermann et al.~\cite{Ackermann11b}). As
the $\gamma$-ray photon spectral index decreases, the high energy
component shifts to the higher energy, and the $\gamma$-rays in
Fermi energy range becomes closer to the high energy peak of SED
for FSRQs. So the negative correlation of the $\gamma$-ray photon
spectral index with the $\gamma$-ray variability index for FSRQs
indicates that the variability amplitude tends to be larger as the
$\gamma$-rays are closer to the high energy peak of SED.

The correlations of radio and $\gamma$-ray luminosity with
$\gamma$-ray photon spectral index for BL Lacs reflect the blazar
sequence among different subclass of BL Lacs (Fossati et
al.~\cite{Fossati98}). For FSRQs, the negative correlation of
$\gamma$-ray luminosity with $\gamma$-ray photon spectral index weas also found in Ackermann et al. (~\cite{Ackermann11b}), where the correlation was very weak. In our results, the correlations become significant when the energy is larger than 1GeV. These correlations
would emerge if the more powerful sources have harder electron spectrums. However, the fast falling above the high energy peak of SED could make the integral luminosity of $\gamma$-ray smaller for the sources with the steeper $\gamma$-ray spectrum, even that the peak luminosity or the total luminosity is higher. This is consistent with the trend that the higher energy bands have better correlations (see Table 1). Our results, which seem opposite to the expectation of the blazar sequence, are caused by this reason with high possibility. So as mentioned in Chen \& Bai (\cite{Chen11}), the $\gamma$-ray luminosity in Fermi range can not be used simply to investigate the blazar sequence instead of the the peak luminosity or the total luminosity, at least for FSRQs. The same trend regarded as harder when brighter is always observed in single sources, e.g., for 3C454.3 (Abdo et al. \cite{Abdo11a}).

In summary, the radio and $\gamma$-ray emission are in good
connection in Fermi/LAT blazars, but these correlations become
worse as the $\gamma$-ray energy increases. Moreover, the flux
correlations would be affected strongly by the selection effects.
The $\gamma$-ray variability index is correlated with other
parameters for FSRQs. These correlations suggest that the $\gamma$-ray
variability may be due to changes inside the $\gamma$-ray emission region like the injected power, rather than changes in the photon density of the external radiation field, and the variability amplitude tends to be larger as the $\gamma$-rays are closer to the high energy peak of SED. The different variability behaviors below and
above the peak energy may cause the different trends of BL Lacs
and FSRQs. The negative correlations of the $\gamma$-ray luminosity with $\gamma$-ray photon spectral index suggest that the $\gamma$-ray luminosity in Fermi range can not be used simply to investigate the blazar sequence instead of the the peak luminosity or the total luminosity, at least for FSRQs. However, our results are also limited by the source
number of our sample, especially for BL Lacs. These results will
be tested with larger samples.

\begin{acknowledgements}
We thank the anonymous referee for insightful comments and constructive suggestions. We are grateful to Yibo Wang, Jiancheng Wang, Deliang
Wang and Zunli Yuan for useful discussions. This research has made
use of data from the MOJAVE database that is maintained by the
MOJAVE team. We thank the National Natural Science Foundation of
China (NSFC; Grant 10903025, 10973034, 11103060, 11133006) for financial
support, and the support of the 973 Program (Grant 2009CB824800).
\end{acknowledgements}

\label{lastpage}

\begin{landscape}
\begin{table}
\begin{center}
\caption[]{The Results of Correlation Test. $\rho$ is the correlation coefficient of Spearman test, r is the correlation coefficient of partial correlation analysis, and P is the chance probability.}
\label{Tab:publ-works}
\scalebox{0.7}{%
 \begin{tabular}{cccccccccccccccccccc}
  \hline
  \multirow{2}*{\centering Source Class} & \multirow{2}*{\centering Energy Band} & \multicolumn{2}{c}{$F_{R}$ VS $F_{\gamma}$} & \multicolumn{2}{c}{$L_{R}$ VS $L_{\gamma}$} & \multicolumn{2}{c}{$L_{R}$ VS $\Gamma$} & \multicolumn{2}{c}{$L_{\gamma}$ VS $\Gamma$} & \multicolumn{2}{c}{$L_{\gamma}/L_{R}$ VS $\Gamma$} & \multicolumn{2}{c}{$L_{R}$ VS $TS_{var}$} & \multicolumn{2}{c}{$L_{\gamma}$ VS $TS_{var}$} & \multicolumn{2}{c}{$L_{\gamma}/L_{R}$ VS $TS_{var}$} &  \multicolumn{2}{c}{$\Gamma$ VS $TS_{var}$}\\
  & & \multicolumn{1}{c}{$\rho$} & \multicolumn{1}{c}{P} & \multicolumn{1}{c}{r} & \multicolumn{1}{c}{P} & \multicolumn{1}{c}{$\rho$} & \multicolumn{1}{c}{P} & \multicolumn{1}{c}{$\rho$} & \multicolumn{1}{c}{P} & \multicolumn{1}{c}{$\rho$} & \multicolumn{1}{c}{P} & \multicolumn{1}{c}{$\rho$} & \multicolumn{1}{c}{P} & \multicolumn{1}{c}{$\rho$} & \multicolumn{1}{c}{P} & \multicolumn{1}{c}{$\rho$} & \multicolumn{1}{c}{P} & \multicolumn{1}{c}{$\rho$} & \multicolumn{1}{c}{P}\\
  \hline
     \multirow{7}*{\centering FSRQ} &
     0.1$<$E$<$100 & 0.38 & $1.1\times10^{-5}$ & 0.39 & $<$0.001 & \multirow{7}*{0.08} & \multirow{7}*{0.36} & -0.26 & 0.003 & -0.47 & $3\times10^{-8}$ & \multirow{7}*{-0.03} & \multirow{7}*{0.72} & 0.35 & $5.9\times10^{-5}$ & 0.59 & $3.9\times10^{-13}$ & \multirow{7}*{-0.47} & \multirow{7}*{$2.8\times10^{-8}$}\\
     & 0.1$<$E$<$0.3 & 0.4 & $2.7\times10^{-6}$ & 0.51 & $<$0.001 & & & -0.03 & 0.7 & -0.19 & 0.03 & & & 0.26 & 0.003 & 0.49 & $6.3\times10^{-9}$ & & \\
     & 0.3$<$E$<$1 & 0.32 & $2\times10^{-4}$ & 0.42 & $<$0.001 & & & -0.18 & 0.05 & -0.38 & $1.3\times10^{-5}$ & & & 0.34 & $7.6\times10^{-5}$ & 0.58 & $8.8\times10^{-13}$ & & \\
     & 0.1$<$E$<$1 & 0.39 & $6.5\times10^{-6}$ & 0.48 & $<$0.001 & & & -0.1 & 0.28 & -0.28 & 0.001 & & & 0.3 & $7\times10^{-4}$ & 0.52 & $2.9\times10^{-10}$ & & \\
     & 1$<$E$<$3 & 0.23 & 0.01 & 0.36 & $<$0.001 & & & -0.3 & $8\times10^{-4}$ & -0.5 & $2.7\times10^{-9}$ & & & 0.38 & $9.8\times10^{-6}$ & 0.6 & $8\times10^{-14}$ & & \\
     & 3$<$E$<$10 & 0.16 & 0.08 & 0.32 & $<$0.001 & & & -0.37 & $2.1\times10^{-5}$ & -0.55 & $1.9\times10^{-11}$ & & & 0.36 & $2.7\times10^{-5}$ & 0.55 & $2.2\times10^{-11}$ & & \\
     & 1$<$E$<$100 & 0.21 & 0.02 & 0.28 & 0.001 & & & -0.43 & $4.1\times10^{-7}$ & -0.61 & $2\times10^{-14}$ & & & 0.44 & $2.8\times10^{-7}$ & 0.63 & $1.9\times10^{-15}$ & & \\
  \hline
     \multirow{7}*{\centering BL Lac} &
     0.1$<$E$<$100 & 0.44 & 0.006 & 0.58 & $<$0.001 & \multirow{7}*{0.66} & \multirow{7}*{$4.2\times10^{-6}$} & 0.42 & 0.008 & -0.72 & $2.5\times10^{-7}$ & \multirow{7}*{0.02} & \multirow{7}*{0.89} & 0.09 & 0.57 & 0.2 & 0.23 & \multirow{7}*{-0.18} & \multirow{7}*{0.27}\\
     & 0.1$<$E$<$0.3 & 0.55 & $3\times10^{-4}$ & 0.78 & $<$0.001 & & & 0.55 & $3\times10^{-4}$ & -0.46 & 0.003 & & & 0.05 & 0.78 & 0.15 & 0.36 & & \\
     & 0.3$<$E$<$1 & 0.34 & 0.03 & 0.68 & $<$0.001 & & & 0.51 & 0.001 & -0.52 & $7\times10^{-4}$ & & & 0.11 & 0.51 & 0.2 & 0.21 & & \\
     & 0.1$<$E$<$1 & 0.49 & 0.002 & 0.74 & $<$0.001 & & & 0.53 & $6\times10^{-4}$ & -0.54 & $5\times10^{-4}$ & & & 0.08 & 0.65 & 0.19 & 0.24 & & \\
     & 1$<$E$<$3 & 0.1 & 0.55 & 0.56 & $<$0.001 & & & 0.43 & 0.006 & -0.64 & $1\times10^{-5}$ & & & 0.14 & 0.4 & 0.28 & 0.09 & & \\
     & 3$<$E$<$10 & -0.11 & 0.49 & 0.51 & 0.001 & & & 0.41 & 0.009 & -0.73 & $1.32\times10^{-7}$ & & & 0.12 & 0.47 & 0.17 & 0.29 & & \\
     & 1$<$E$<$100 &0.04 & 0.8 & 0.43 & 0.007 & & & 0.35 & 0.03 & -0.77 & $7.6\times10^{-9}$ & & & 0.15 & 0.38 & 0.26 & 0.12 & & \\
  \hline
     \multirow{7}*{\centering All Sources} &
     0.1$<$E$<$100 & 0.47 & $1.2\times10^{-10}$ & 0.64 & $<$0.001 & \multirow{7}*{0.34} & \multirow{7}*{$6.2\times10^{-6}$} & 0.1 & 0.21 & -0.43 & $6.2\times10^{-9}$ & \multirow{7}*{0.08} & \multirow{7}*{0.32} & 0.37 & $1.2\times10^{-6}$ & 0.5 & $7\times10^{-12}$ & \multirow{7}*{0.28} & \multirow{7}*{$2\times10^{-4}$}\\
     & 0.1$<$E$<$0.3 & 0.51 & $4.1\times10^{-12}$ & 0.76 & $<$0.001 & & & 0.27 & $4\times10^{-4}$ & -0.09 & 0.23 & & & 0.3 & $7.6\times10^{-5}$ & 0.45 & $1.3\times10^{-9}$ & & \\
     & 0.3$<$E$<$1 & 0.41 & $5.4\times10^{-8}$ & 0.69 & $<$0.001 & & & 0.17 & 0.03 & -0.28 & $3\times10^{-4}$ & & & 0.37 & $1.3\times10^{-6}$ & 0.51 & $1.4\times10^{-12}$ & & \\
     & 0.1$<$E$<$1 & 0.49 & $3.1\times10^{-11}$ & 0.74 & $<$0.001 & & & 0.23 & 0.004 & -0.19 & 0.01 & & & 0.33 & $1.6\times10^{-5}$ & 0.47 & $1.3\times10^{-10}$ & & \\
     & 1$<$E$<$3 & 0.26 & $8\times10^{-4}$ & 0.61 & $<$0.001 & & & 0.07 & 0.35 & -0.45 & $1.3\times10^{-9}$ & & & 0.4 & $1.1\times10^{-7}$ & 0.52 & $6.4\times10^{-13}$ & & \\
     & 3$<$E$<$10 & 0.11 & 0.18 & 0.56 & $<$0.001 & & & 0.02 & 0.85 & -0.56 & $8\times10^{-15}$ & & & 0.37 & $7.1\times10^{-7}$ & 0.43 & $7\times10^{-9}$ & & \\
     & 1$<$E$<$100 & 0.22 & 0.004 & 0.5 & $<$0.001 & & & -0.04 & 0.6 & -0.6 & $1.3\times10^{-17}$ & & & 0.44 & $3.7\times10^{-9}$ & 0.52 & $4.9\times10^{-13}$ & & \\
  \hline
\end{tabular}}
\end{center}
\end{table}
\end{landscape}

\end{document}